\documentclass[pdflatex,sn-aps]{sn-jnl}

\usepackage[utf8]{inputenc}
\usepackage{graphicx}%
\usepackage{pdfpages}
\usepackage{multirow}%
\usepackage{amsmath,amssymb,amsfonts}%
\usepackage{amsthm}%
\usepackage{mathrsfs}%
\usepackage[title]{appendix}%
\usepackage{xcolor}%
\usepackage{textcomp}%
\usepackage{manyfoot}%
\usepackage{booktabs}%
\usepackage{algorithm}%
\usepackage{algorithmicx}%
\usepackage{algpseudocode}%
\usepackage{listings}%

\theoremstyle{thmstyleone}%
\theoremstyle{thmstyletwo}%

\theoremstyle{thmstylethree}%
\usepackage{ragged2e}
\usepackage{caption}
\begin{document}

\title[Zero-field plasmon splitter]{Zero-field plasmon splitter: Controlled interaction between quantum anomalous Hall edge states}


\author[1]{\fnm{Dominik} \sur{Hoborka}}

\author[1]{\fnm{Torsten} \sur{Röper}} 

\author[2]{\fnm{Fabian} \sur{Hassler}}

\author[1]{\fnm{Alexey A.} \sur{Taskin}}

\author[1]{\fnm{Yoichi} \sur{Ando}}

\author*[1]{\fnm{Erwann} \sur{Bocquillon}
}\email{bocquillon@ph2.uni-koeln.de}
\affil[1]{
\orgdiv{II. Physikalisches Institut},
\orgname{Universität zu Köln},
\orgaddress{
\city{Köln},
\postcode{50937},
\country{Germany}}}

\affil[2]{
\orgdiv{Institute for Quantum Information},
\orgname{RWTH Aachen University},
\orgaddress{
\city{Aachen},
\postcode{52056},
\country{Germany}}}

\abstract{
The quantum anomalous Hall (QAH) effect provides chiral edge states without the need for an external magnetic field, but unlike in conventional quantum Hall systems, their trajectories cannot be readily controlled electrostatically. Here, we engineer Coulomb interactions between two spatially separated, counter-propagating QAH edge channels using a floating metallic island. Microwave measurements reveal frequency-dependent redistribution of injected edge plasmons between two output channels, demonstrating interaction-induced hybridization of the collective modes. A long-range interaction model quantitatively reproduces the response and reveals a very low plasmon velocity in the coupling region, strongly reduced by electrostatic screening. Our results establish engineered Coulomb coupling as a route to control collective excitations in QAH edge states and provide a building block for chiral plasmonic circuits and electron quantum optics in magnetic topological insulators.
}

\maketitle
\newpage
\section*{Introduction}
The quantum anomalous Hall (QAH) effect in magnetic topological insulators supports chiral edge transport without the need for an external magnetic field~\cite{chang_experimental_2013,chang_high-precision_2015,yu_quantized_2010}. These edge states also support collective charge excitations, or edge plasmons, which can be excited and detected at microwave frequencies~\cite{roper_propagation_2024,mahoney_zero-field_2017,martinez_edge_2023}. Their propagation direction is set by the magnetization of the material and can be reversed by magnetic-field or current-induced switching~\cite{rupp_dynamics_2026,yuan_electrical_2024,zimmermann_current-induced_2024}. This makes QAH systems interesting for applications ranging from chiral electronics and quantum metrology to quantum-information concepts~\cite{viola_hall_2014,mahoney_-chip_2017,rodenbach_metrological_2022,huang_quantum_2025,beenakker_deterministic_2019}.
Controlling how two QAH edge states couple is an important step toward using them in more complex devices. In one-dimensional electronic systems, Coulomb interactions lead to collective charge modes whose propagation depends strongly on the surrounding electrostatic environment~\cite{giamarchi_quantum_2003}. In conventional quantum Hall systems, capacitive coupling between nearby edge channels has been used to study plasmon propagation, mode coupling, and charge fractionalization~\cite{kumada_edge_2011,kamata_voltage-controlled_2010,bocquillon_separation_2013,levkivskyi_energy_2012,berg_fractional_2009}. Importantly, such coupling allows the charge dynamics of one edge channel to influence another without requiring electrons to tunnel between them.\\
Quantum Hall systems also offer direct control over chiral edge trajectories through quantum point contacts. These can partition electronic excitations between different paths and have been used to realize electronic beam splitters and interferometers~\cite{bocquillon_coherence_2013,glidic_quasiparticle_2023,frigerio_gate_2024,kouame_electrostatic_2026}. In a QAH insulator, however, there is only a single chiral edge state, and an equivalent electrostatic control of its path is not readily available. An alternative approach is therefore required
to manipulate QAH edge excitations without relying on electron tunneling.\\
Coulomb interactions provide such an alternative. When two spatially separated edge states are brought close together, their collective charge excitations can interact through their electrostatic environment even in the absence of particle exchange. Such interaction-mediated coupling has been extensively studied in conventional quantum Hall edge channels~\cite{bocquillon_separation_2013,levkivskyi_energy_2012,hashisaka_distributed-element_2013,kamata_fractionalized_2014,perfetto_time-resolved_2014}, where it can lead to hybridization of co- or counter-propagating plasmons and to frequency-dependent redistribution of an injected excitation between different channels. Implementing and controlling this mechanism in the QAH regime would provide a new route toward manipulating chiral collective excitations at zero magnetic field.\\
Here, we engineer such an interaction between two spatially separated, counter-propagating QAH edge states using a floating metallic island. The island provides a common electrostatic environment for the two edges and thereby mediates a nonlocal Coulomb coupling. We realize this geometry as a plasmon splitter in which an excitation injected into one QAH edge can emerge in either of two spatially separated output channels. Microwave transmission measurements reveal frequency-dependent oscillations in the relative transmission between the two output paths, demonstrating interaction-induced hybridization and mode conversion of the coupled edge plasmons.\\
The coupling region strongly modifies the plasmon propagation through electrostatic screening. The metallic island screens the electric field associated with charge excitations along the edge, thereby increasing the effective capacitance per unit length and equivalently reducing the plasmon velocity. By comparing devices with different interaction-region lengths and with a reference device without the metallic island, we extract the propagation time and the effective plasmon velocity in the coupled region. We find a velocity of $v_{\rm s}=(12\pm4)$~km\,s$^{-1}$, more than an order of magnitude smaller than the velocities $v_{\rm s}=(300-500)$~km\,s$^{-1}$ reported for isolated QAH edge states~\cite{roper_propagation_2024, mahoney_zero-field_2017,martinez_edge_2024,roper_breakdown_2025}. The resulting shorter plasmon wavelength brings the interaction-induced mode conversion into the experimentally accessible microwave-frequency range.
Our results establish nonlocal Coulomb coupling as a means of engineering interactions between QAH edge states without modifying their topology or introducing electron tunneling. More generally, the demonstrated control of collective excitations provides a route toward interaction-engineered chiral plasmonic circuits and interferometric devices operating at zero magnetic field.

\section*{Results}
\subsection*{Principles of the experiment}
\begin{figure}[h!]
        \centering
        \includegraphics[width=1\textwidth,keepaspectratio=true]{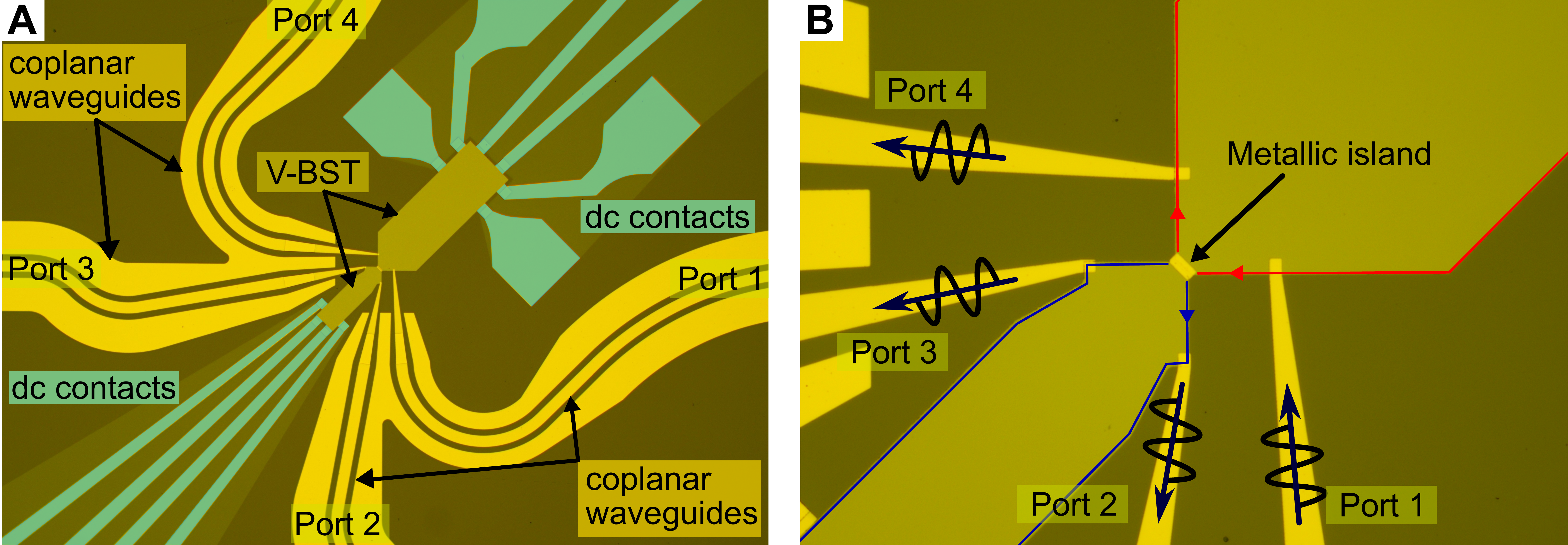}
   
        
        \caption{\textbf{Plasmon splitter device.} \textbf{(A)} Full microscope image of a splitter device. Ohmic contacts (highlighted in light blue) allow for Hall measurements and coplanar waveguides transmit microwave signals to the four ports used for microwave experiments. \textbf{(B)} Zoomed-in image of a device. The two V-BST bars (light brown) are physically separated by a thin trench in the center. The edge states of each bar (highlighted by the red and blue lines) are capacitively coupled via the metallic island covering the trench. Port 1 is used for the microwave excitation, and the other three for detection of the transmitted signal, yielding the scattering parameters $S_{21}$, $S_{31}$ and $S_{41}$.}
        \label{fig:sample}
\end{figure}

\begin{figure}[h]
        \centering

        \includegraphics[width=1\textwidth,keepaspectratio=true]{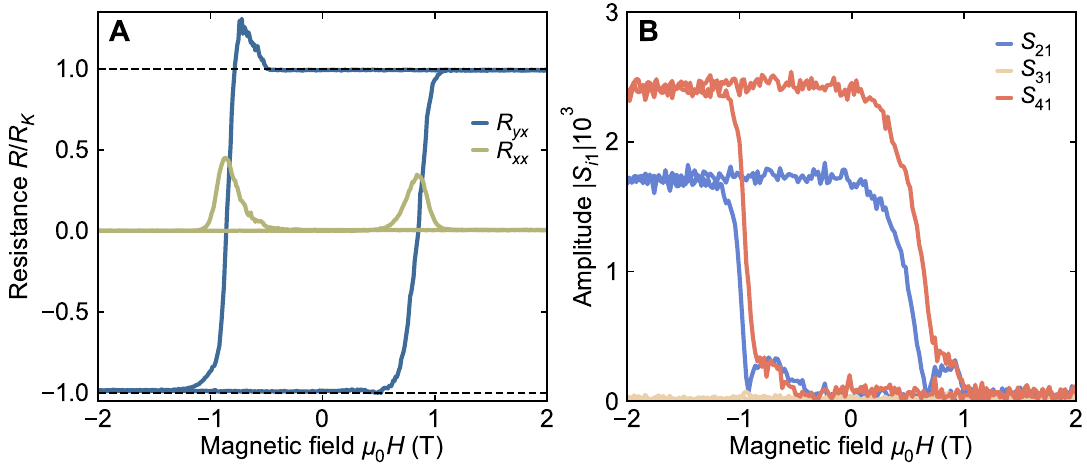}
        \caption {\textbf{Magnetic field dependence of dc and microwave response.} \textbf{(A)} Magnetic field sweeps of dc longitudinal and transverse resistances $R_{\rm xx}$ (light green lines) and $R_{\rm yx}$ (violet lines). The black dashed lines correspond to $R=\pm R_{\rm K}$, with $R_{\rm K}=h/e^2$. \textbf{(B)} Magnetic field sweep of corrected transmission amplitude of $S_{21}$ (blue lines), $S_{31}$ (beige lines) and $S_{41}$ (red lines) for $f=1.5$~GHz and an excitation power of $P\approx -51$~dBm (VNA output power of $-10$~dBm). Both measurements are taken on device A at $T\approx 20$~mK.}
        \label{fig:fieldsweep}
\end{figure}

We study the Coulomb interaction between chiral edge states of the QAH insulator V-doped (Bi$_{1-x}$Sb$_x$)$_2$Te$_3$ (V-BST)~\cite{chang_experimental_2013,chang_high-precision_2015} by fabricating a plasmon splitter device, as shown in Fig.~\ref{fig:sample}.
\newline
The plasmon splitter device comprises a coupling region in which two counter-propagating edge channels coexist in close proximity. 
This is achieved by etching a thin trench into the V-BST layer, separating it into two disconnected bars. A floating metallic island is placed above the trench by capacitively coupling it via a 14~nm-thick $\text{Al}_2\text{O}_3$ layer serving as a dielectric. The metallic island strongly enhances the inter-edge coupling within the coupling region. Narrow microwave finger gates, connected to coplanar waveguides (labeled Port 1-4), excite and detect edge plasmons via a four-port vector network analyzer (VNA). Port 1 injects the microwave signal, while the remaining three ports measure the transmission ($S_{21}$, $S_{31}$, $S_{41}$) through the edge channels. Ohmic contacts (located away from the island) allow for the dc device characterization via standard low-frequency lock-in measurements, confirming the QAH state. Device A exhibits a Hall resistance within 2\% of $R_{\rm K}$ with small longitudinal resistance as shown in Fig.~\ref{fig:fieldsweep}A. The dc characteristics of all devices are summarized in Table~\ref{tab:Device-list}. Using the ohmic contacts, we verify electrically that the island is capacitively coupled to the V-BST bars and is not ohmically connected.\newline\newline We investigated six splitter devices with different dimensions of the metallic island. Two devices contain no metallic island to quantify the residual inter-edge coupling. The remaining four devices incorporate metallic islands with trench lengths $L$ of 2~µm, 6~µm (two devices), and 12~µm, enabling a systematic study of the length dependence.
We focus on three devices in the main text. The lengths of the metallic islands on devices A and B are 6~µm and 12~µm, respectively (see Fig.~\ref{fig:sample}). Device C has no metallic island and a trench length of 2~µm. We discuss the other devices in Sec.~S7 of the Supplementary Material.

\subsection*{Evidence for inter-edge coupling}
\begin{figure}[t]
    \centering
    \includegraphics[width=1\textwidth,keepaspectratio=true]{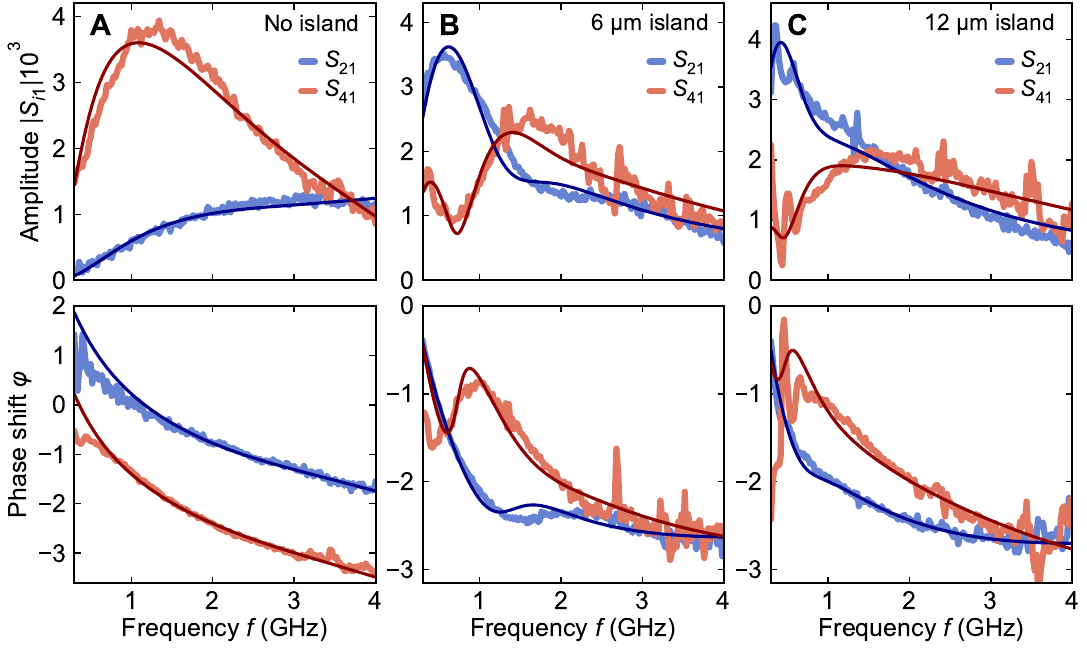}
    \caption{
    \textbf{Microwave transmission $S_{21}$ and $S_{41}$.} Amplitude and phase of $S_{21}$ and $S_{41}$ for \textbf{(A)} Device C without a metallic island, \textbf{(B)} Device A with a 6~µm-long metallic island, and \textbf{(C)} Device B with a 12~µm-long metallic island. The solid lines show fits using the complete transmission model of Eq.~\eqref{eq:stotal}. The interaction parameters $\tau_{\rm p,s}$ and $\tau_{\rm r,s}$ are fixed to the values extracted from the ratio analysis in Fig.~\ref{fig:fits}, while the remaining parameters $\tau_{\rm p,e}$ and $\tau_{\rm r,e}$ are determined in a second step, as detailed in the main text. The fit parameters are shown in Tab.~\ref{tab:fit_parameters}. Measurements are taken for $\mu_0H=-2$~T and $T\approx 20$~mK.}
        \label{fig:fit_s_total}
\end{figure}

We first establish inter-edge coupling and show that the metallic island strongly enhances the interaction between the two edge channels. Fig.~\ref{fig:fieldsweep}B displays the magnetic-field sweeps of the scattering-parameter amplitudes for each detection port, recorded at the frequency $f=1.5$~GHz.
In the present geometry, a negative magnetization (at -~2~T) results in clockwise chirality, as highlighted in Fig.~\ref{fig:sample}B. In that case, plasmons injected in port 1 are directed towards the coupling region and are detected at ports 2 and 4, yielding a finite transmission $S_{21}$ and $S_{41}$. A positive magnetization of the sample (at +2~T) corresponds to a counterclockwise chirality of the edge channel, resulting in no transmission (see Materials and Methods). In analogy to the dc characterization, the microwave signals measured at ports 2 and 4 exhibit a pronounced hysteresis upon reversal of the magnetization. The hysteretic microwave response confirms that the detected signals originate from chiral QAH edge transport~\cite{roper_propagation_2024}. For clockwise chirality, the simultaneous finite transmission to ports 2 and 4 indicates redistribution of the injected excitation within the coupling region. By contrast, the corrected signal at port 3, upstream, remains small, consistent with the expected propagation direction in this geometry. A weak hysteretic contribution is nevertheless observed at port 3 for some devices.\newline\newline
We now turn to analyzing the frequency-dependent amplitude and phase of both transmission coefficients $S_{21}$ and $S_{41}$. The amplitude and phase of the spectra of samples A and B are shown in Fig.~\ref{fig:fit_s_total}B and C, respectively.
Interestingly, at low frequencies, the signal transmitted to the opposite side of the plasmon splitter $S_{21}$ exceeds the signal remaining on the same side $S_{41}$. This behavior reverses at higher frequencies (around 1.2~GHz for device A and around 1.6~GHz for device B), such that $S_{41}$ becomes the dominant transmission channel. Both transmission amplitudes exhibit oscillations before approaching the same value at higher frequencies. The oscillations are more pronounced for Device A (6~µm-long splitter) than for Device B (12~µm-long splitter), which we attribute to stronger dissipation in the longer island, leading to equal amplitudes in $S_{21}$ and $S_{41}$ already above 1.6~GHz.
These oscillations are absent in reference device C without a metallic island, as shown in Fig.~\ref{fig:fit_s_total}A. Here, most of the signal is transmitted via $S_{41}$, while $S_{21}$ increases slowly with frequency. This demonstrates that, in the absence of a metallic island, a weak residual coupling remains, whereas the metallic island mediates a much stronger coupling between the edge states, leading to an oscillatory behavior, previously observed for co-propagating quantum Hall edge channels~\cite{bocquillon_separation_2013}.\newline\newline
The coupling between the two edge channels depends on the phase, which the injected plasmon accumulates while traversing the coupling region. For an accumulated phase $\phi=2\pi n$, the positive and negative charge oscillations cancel over the length of the coupling region, leaving the metallic island uncharged and resulting in only weak capacitive coupling between the edge channels and maximal transmission in $S_{41}$. In contrast, for $\phi=(2n+1)\pi$, the oscillations add constructively, leading to maximal charging of the floating island and therefore to the strongest coupling between the counter-propagating edge channels. As the microwave frequency increases, the accumulated phase changes continuously, giving rise to the oscillatory splitting observed experimentally. Electrostatic screening by the metallic island strongly reduces the plasmon velocity and therefore shortens the plasmon wavelength compared to the bare edge, bringing the oscillation to an observable range $f<4$~GHz. Based on this physical picture, we develop a quantitative model of the plasmon splitter in the following.
\subsection*{Long-range interaction model}
\begin{figure}[h!]
        \centering
        \includegraphics[width=0.5\textwidth,keepaspectratio=true]{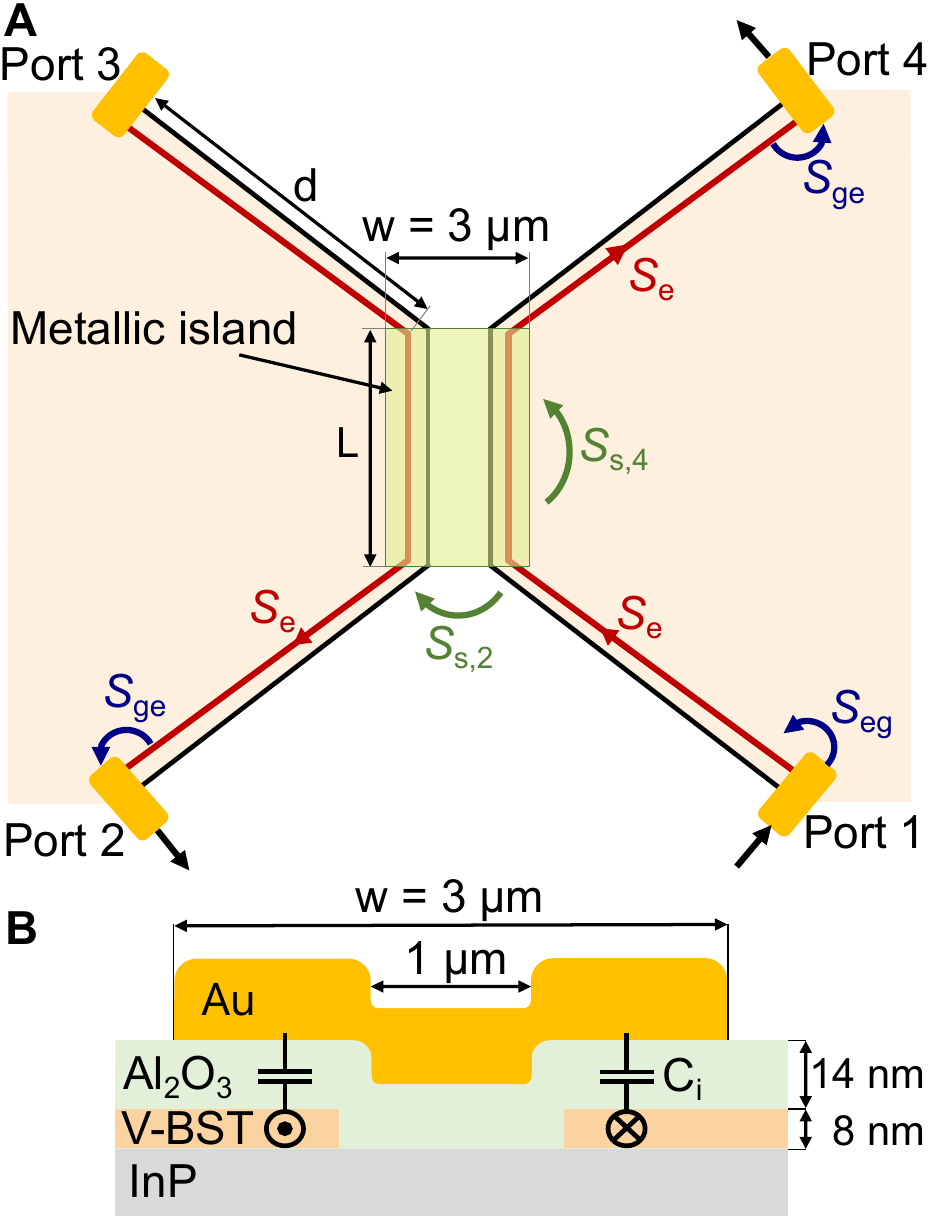}   
        \caption{\textbf{Model of the plasmon splitter.} \textbf{(A)} Equivalent circuit describing the microwave transmission. The measured scattering parameters are decomposed into three contributions: microwave excitation $S_{\rm eg}$ and detection $S_{\rm ge}$, propagation along the bare edge channels $S_{\rm e}$, and propagation through the plasmon splitter $S_{\rm s,2}$ and $S_{\rm s,4}$. The bare edge is characterized by the propagation and relaxation times $\tau_{\rm p,e},\tau_{\rm r,e}$, while the splitter is described by $\tau_{\rm p,s},\tau_{\rm r,s},\tau_{\rm c,s}$. \textbf{(B)} Cross-sectional schematic of the plasmon splitter. Two counter-propagating edge channels form collective plasmon modes mediated by a floating metallic island connected via a capacitance $C_{\rm i}$, corresponding to the charging time $\tau_{\rm c,s}=R_{\rm K}C_{\rm i}$.}
        \label{fig:TransmissionLine}
\end{figure}

To quantitatively describe the measured microwave transmission, we model the device using a distributed-circuit approach that has previously been shown to accurately capture the excitation and propagation of edge plasmons in both quantum Hall and QAH systems~\cite{kamata_fractionalized_2014,gourmelon_characterization_2020,roper_propagation_2024,hashisaka_distributed-element_2013,bocquillon_separation_2013}. Here, we extend this framework to include the capacitive coupling between two counter-propagating edge states mediated by the floating metallic island. As illustrated in Fig.~\ref{fig:TransmissionLine}A, the device is described by three scattering coefficients corresponding to (i) microwave excitation $S_{\rm eg}$ and detection $S_{\rm ge}$, (ii) propagation along the bare edge channels $S_{\rm e}$, and (iii) the plasmon splitter $S_{{\rm s},i}$.
Microwave excitation and detection occur capacitively through the gate capacitance $C_{\rm g}$ and are modeled following Ref.~\cite{gourmelon_characterization_2020}. Propagation along the bare edge channels follows the distributed transmission-line model developed in Ref.~\cite{roper_propagation_2024} and takes charge puddles screening into account. We introduce a propagation time $\tau_{\rm p,e}$ and a relaxation time $\tau_{\rm r,e}$, yielding the complex wavevector $k_{\rm e}(\omega)$ such that:
\begin{equation}
k_{\rm e}\,d=\frac{\omega\tau_{\rm p,e}}{1+i\omega\tau_{\rm r,e}},\label{eq:dispersion}
\end{equation}
Here, $\tau_{\rm p,e}=d/v_{\rm e}$ describes the propagation time through a section of length $d$ at a velocity $v_{\rm e}$, while $\tau_{\rm r,e}$ phenomenologically accounts for frequency-dependent dissipation. Throughout this work, Eq.~\eqref{eq:dispersion} is also used to describe propagation in the coupling region, with different values of $\tau_{\rm p,s}$ and $\tau_{\rm r,s}$ reflecting the different electrostatic environments.
For the bare edge channels, the propagation is described by the scattering coefficient $S_{\rm e}=A\,\exp(-ik_{\rm e}d)$, where $k_{\rm e}$ follows Eq.~\eqref{eq:dispersion} with the parameters $\tau_{\rm p,e}$ and $\tau_{\rm r,e}$ and $A$ is a real number that accounts for losses in dc.\newline\newline
The coupling region requires an additional description because the two counter-propagating edge states become capacitively coupled underneath the metallic island. Two limiting cases have previously been proposed~\cite{bocquillon_separation_2013}. The zero-range model assumes local capacitive coupling along the propagation direction in a Luttinger liquid framework~\cite{lee_spin-charge_1997,sukhorukov_resonant_2007}, whereas the long-range model describes the interaction through the common electrostatic potential of a floating conductor represented by a lumped capacitance $C_{\rm i}$.
In our device, the metallic island remains at an equipotential because its internal RC time is much shorter than the plasmon propagation times of several hundred picoseconds. The long-range model is therefore expected to apply. A plasmon propagating along either edge therefore modifies the electrostatic potential experienced by both edge channels over the entire length of the island, mediating a nonlocal Coulomb interaction between them. The effective interaction strength is consequently set by the island length $L$, rather than by the distance of the edge channels. Solving the coupled equations for the collective plasmon modes (see Sec.~S1 of the Supplementary Material for details) yields:
\begin{equation}
\label{eq:ratio}
S_{\rm s,2}=\frac{1-e^{-ik_{\rm s}L}}{2\biggl(1+\dfrac{i}{\omega\tau_{\rm c,s}}\bigl(1-e^{-ik_{\rm s}L}\bigr)\biggr)},\qquad
S_{\rm s,4}=1-S_{\rm s,2},
\end{equation}
where $\tau_{\rm c,s}=R_{\rm K}C_{\rm i}$ is the charging time of the floating metallic island acting as a splitter. The propagation inside the coupling region is described by the same dispersion relation, Eq.~\eqref{eq:dispersion}, using different parameters $\tau_{\rm p,s}$, $\tau_{\rm r,s}$ and replacing the bare edge length $d$ with the length $L$ of the coupling region.\newline\newline
To reduce the number of fit parameters, we consider the appropriate limiting case for each device. For devices containing a metallic island, the investigated frequencies satisfy $\omega \tau_{\rm c,s}\gg1$ and the charging dynamics of the floating island can be neglected such that the response depends only on the two parameters $\tau_{\rm p,s}$ and $\tau_{\rm r,s}$. For the reference device without a metallic island, we instead set $\tau_{\rm r,s}=0$, corresponding to the absence of additional dissipation in the coupling region, while retaining $C_{\rm i}$ to account for the residual inter-edge coupling capacitance $C_{\rm s}=C_{\rm i}/2$. Finally, owing to the chiral nature of the edge channels and the absence of reflections, the total transmission is obtained as
\begin{equation}\label{eq:stotal}
    S_{i1}=S_{\rm ge} S_{\rm e} S_{{\rm s},i}S_{\rm e}S_{\rm eg}, \quad i=2,4.
\end{equation}
Although the model contains only a small number of physically meaningful parameters, several of them contribute simultaneously to the measured transmission spectra. To reduce parameter correlations, we first determine the splitter parameters $\tau_{\rm p,s}$ and $\tau_{\rm r,s}$ from the ratio $\eta = S_{21}/S_{41}$ (denoted hereafter as the splitting ratio). The scattering coefficient for the common excitation port $S_{\rm eg}$ and bare-edge propagation $S_{\rm e}$ are expected to cancel out for $S_{21}$ and $S_{41}$, thanks to the symmetric device geometry. The parameters $\tau_{\rm p,s}$ and $\tau_{\rm r,s}$ are subsequently fixed when fitting the complete transmission according to Eq.~\eqref{eq:stotal}. A comparison with the zero-range interaction model is presented in Sec.~S2 of the Supplementary Material.

\subsection*{Comparison between model and data}
\begin{figure}[ht]
        \centering
        \includegraphics[width=1\textwidth,keepaspectratio=true]{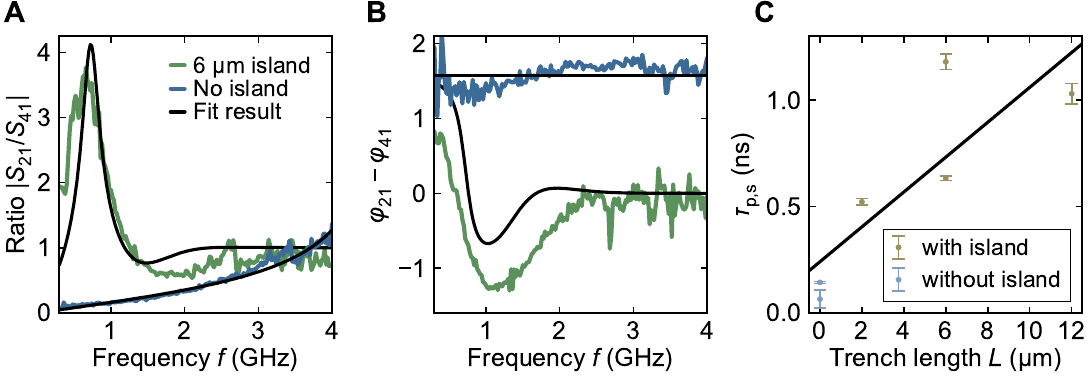}
        \caption{\textbf{Frequency dependence of the splitting ratio $S_{21}$/$S_{41}$.} \textbf{(A)} Amplitude and \textbf{(B)} phase of the ratio as a function of the frequency for Devices A (6~µm metallic island, shown as green line) and C (no metallic island, shown as violet line). The solid lines represent the fit with Eq.~\eqref{eq:ratio}. We find: $\tau_{\rm p,s}=(630\pm10)$~ps, $\tau_{\rm r,s}=(31\pm2)$~ps for Device A and $\tau_{\rm p,s}=(144\pm5)$~ps, $C_{\rm i}=(3.0\pm0.1)$~fF for Device C. \textbf{(C)} The fit parameter $\tau_{\rm p,s}$ is shown as a function of the island length $L$. Light green points correspond to Devices A, B, D, E, and light blue points to the reference Devices C and F, which are plotted at $L=0$ to account for the residual inter-edge coupling without a metallic island. The fit parameter $\tau_{\rm p,s}$ stems from fits of the ratio $S_{21}/S_{41}$, as shown in \textbf{(A)} and \textbf{(B)}. The error bars show the fit uncertainties. The black line corresponds to a linear fit, which yields a velocity of $v_{\rm s}=(12\pm4)$~km\,s$^{-1}$.  All measurements are taken for $\mu_0H=-2$~T and $T\approx 20$~mK.}
        \label{fig:fits}
\end{figure}

The solid lines in Fig.~\ref{fig:fits}A,B show fits using the simplified long-range interaction model in Eq.~\eqref{eq:ratio} describing the amplitude and phase of the splitting ratio $\eta$. Despite containing only the two free parameters $\tau_{\rm p,s}$ and $\tau_{\rm r,s}$, the model reproduces the oscillation period and the overall frequency dependence over the complete experimental bandwidth. The only noticeable discrepancy is a systematic overestimation of the phase. As shown in Sec.~S3 of the Supplementary Material, this originates from unequal gate-edge coupling $S_{ge}$ of the two detection gates.
\newline\newline
For Device~A (6~µm-long metallic island), the fit yields a propagation time of $\tau_{\rm p,s}=(630\pm10)$~ps and a relaxation time of $\tau_{\rm r,s}=(31\pm2)$~ps. Repeating the analysis for different island lengths yields the propagation times summarized in Fig.~\ref{fig:fits}C, including the propagation times for samples without metallic islands shown for $L=0$. Their roughly linear dependence on the island length $L$ demonstrates that the measured propagation times accumulate within the coupling region. A linear fit gives a propagation velocity of $v_{\rm s}=(12\pm4)$~km\,s$^{-1}$ in the coupling region, about 20 to 50 times smaller than in bare edge states~\cite{roper_propagation_2024}.
The pronounced reduction of $v_{\rm s}$ is an expected consequence of the electrostatic screening introduced by the floating metallic island, which increases the effective capacitance and thereby slows the collective plasmon mode.
This relation can be quantified using the dispersion relation of Eq.~\eqref{eq:dispersion}. In the experimentally relevant regime $\omega\tau_{\rm r,s}<1$, the dispersion can be approximated as $k_{\rm s}L\approx\omega\tau_{\rm p,s}$, yielding a group velocity $v_{\rm s}=\partial\omega/\partial k_{\rm s}=L/\tau_{\rm p,s}$. Within the distributed-capacitance description, the propagation time is related to the distributed capacitance $\mathcal{C}_{\rm e}$ of the edge channel by $\tau_{\rm p,s}=R_{\rm K}\mathcal{C}_{\rm e}L$. The line capacitance $\mathcal{C}_{\rm e}$ accounts for the electrostatic environment of the edge channel, including bare edge interactions and coupling to charge puddles and the metallic gates~\cite{roper_propagation_2024,kumada_suppression_2020}. As observed on gated devices~\cite{kumada_suppression_2020}, the capacitance to the gate is the relevant quantity, and hence we assume that the geometric capacitance of the metallic island is the dominant contribution to $\mathcal{C}_{\rm e}$. Therefore, we estimate $\mathcal{C}_{\rm e}$ geometrically as $\mathcal{C}_{\rm e}=c_{\rm geo}w$. Here, $c_{\rm geo}$ is the capacitance per unit area between the metallic island and the edge channel, and $w$ is the effective overlap width, which we approximate as $w=1$~µm. The experimentally determined $\tau_{\rm p,s}$ then corresponds to $c_{\rm geo}\approx3$~mF\,m$^{-2}$, in reasonable agreement with the geometrical estimate $c_{\rm geo}=\varepsilon_0\varepsilon_r/t\approx4.4$~mF\,m$^{-2}$ for a $t=14$~nm-thick Al$_2$O$_3$ dielectric layer ($\varepsilon_r\approx7$). For the reference device without a metallic island (Device C), we estimate an effective interaction length $L_{\rm eff}\approx11$~µm to account for the spatial extent of the residual inter-edge coupling (see Sec.~S4 of the Supplementary Material). The resulting effective velocity $v_{\rm s}\approx80$~km\,s$^{-1}$ remains below the $v_{\rm e}\approx(200{-}500)$~km\,s$^{-1}$ reported for isolated QAH edge states~\cite{roper_propagation_2024}. This reduction is consistent with the formation of hybridized collective plasmon modes, whose propagation is modified by Coulomb coupling between the counter-propagating edge channels~\cite{flensberg_coulomb_1998,fu_non-reciprocal_2025}.
\newline\newline
Besides reducing the propagation velocity, the metallic island also introduces relaxation, quantified by the time constant $\tau_{\rm r,s}$. This relaxation limits the coherence of the coupled plasmon modes and suppresses the oscillations above the characteristic frequency
$f_c\simeq\frac{1}{2\pi\sqrt{\tau_{\rm p,s}\tau_{\rm r,s}}}\approx1$~GHz, where the transmission gradually approaches the symmetric limit $S_{21}=S_{41}$, corresponding to a splitting ratio  $\eta=1$ of the incoming plasmon.\newline\newline
Having established the intrinsic properties of the plasmon splitter, we next consider the complete transmission model. The previously determined values of $\tau_{\rm p,s}$ and $\tau_{\rm r,s}$ are kept fixed, while the remaining parameters describing bare-edge propagation and microwave coupling are obtained by fitting the complete transmission according to Eq.~\eqref{eq:stotal}. The excellent quantitative agreement with both amplitude and phase over the full frequency range in Figures~\ref{fig:fit_s_total}A-C provides a stringent test and validates the model. Table~\ref{tab:fit_parameters} summarizes the extracted fit parameters.
\begin{table}[t]
    \centering
    \captionsetup{width=\textwidth}
    \caption{\textbf{Extracted fit parameters from the complete transmission model.} For Device A (6~µm long metallic island), Device B (12~µm long metallic island), and the reference device without a metallic island (Device C). For Devices A and B, $\tau_{\rm p,s}$ and $\tau_{\rm r,s}$ are extracted from fits of the splitting ratio $\eta =S_{21}$/$S_{41}$ with Eq.~\eqref{eq:ratio} in the high-frequency limit $\omega\tau_{\rm c,s}\gg 1$. For Device C, $\tau_{\rm p,s}$ and $C_{\rm i}$ are extracted from fits of $\eta$ with Eq.~\eqref{eq:ratio} in the limit $\tau_{\rm r,s}\rightarrow 0$. The bare-edge parameters $\tau_{\rm p,e}$, $\tau_{\rm r,e}$, and the gate capacitance $C_g$ are then extracted from fits of the full transmission $S_{21}$ and $S_{41}$ with Eq.~\eqref{eq:stotal}, while fixing the splitter parameters $\tau_{\rm p,s}$, $\tau_{\rm r,s}$, and $\tau_{\rm c,s}$ to the values extracted from the corresponding fits of $\eta$.}
    \label{tab:fit_parameters}
    \begin{tabular}{|c|c|c|c|c|c|c|}
        \hline
        Device
        & \multicolumn{2}{c|}{A ($L=6$~µm) }
        & \multicolumn{2}{c|}{B ($L=12$~µm) }
        & \multicolumn{2}{c|}{C (no island) }
        \\
        \hline
        & $S_{21}$ & $S_{41}$ & $S_{21}$ & $S_{41}$ & $S_{21}$ & $S_{41}$ \\
        \hline
        $\tau_{\rm p,s}$ [ps]
        & \multicolumn{2}{c|}{$630\pm 10$}
        & \multicolumn{2}{c|}{$1030\pm 10$}
        & \multicolumn{2}{c|}{$144 \pm 5$}
        \\
        $\tau_{\rm r,s}$ [ps]
        & \multicolumn{2}{c|}{$31\pm 2$}
        & \multicolumn{2}{c|}{$83\pm 14$}
        & \multicolumn{2}{c|}{$-$}
        \\
        $C_{\rm i}$ [fF]
        & \multicolumn{2}{c|}{$-$}
        & \multicolumn{2}{c|}{$-$}
        & \multicolumn{2}{c|}{$3.0 \pm 0.1 $}
        \\
        \hline
        $\tau_{\rm p,e}$ [ps]
        & $43 \pm 1$
        & $36\pm 1$
        & $59 \pm 1$
        & $33\pm 1$
        & $50\pm 2$
        & $46\pm 1$
        \\
        $\tau_{\rm r,e}$ [ps]
        & $39\pm 1$
        & $27\pm 2$
        & $50\pm 1$
        & $18\pm 2$
        & $45\pm 1$
        & $37\pm 1$
        \\
        $C_{\rm g}$ [fF] 
        & $94\pm 3$
        & $15 \pm 1$
        & $42\pm 2$
        & $17 \pm 1$
        & $11.8\pm 0.7$
        & $13.1\pm 0.3$
        \\
        \hline
    \end{tabular}
\end{table}
For all three devices, the bare-edge propagation times $\tau_{\rm p,e}$ are of comparable magnitude, ranging from approximately $30$ to $60$~ps across the two transmission paths, which correspond to bare-edge velocities of $v_{\rm e}\approx(300-600)$~km\,s$^{-1}$, consistent with previous measurements on isolated QAH edge states~\cite{roper_propagation_2024}. This agreement supports the separation of the transmission into propagation along the bare edge and propagation within the coupling region. The extracted relaxation times $\tau_{\rm r,e}$ are likewise comparable across the devices, although some variation between the two transmission paths remains. The fit additionally yields different gate capacitances $C_{\rm g}$ for the individual detection gates, reflecting differences in their coupling to the edge. These asymmetries account for the small residual phase mismatch observed in the simplified ratio analysis (see Fig.~\ref{fig:fits}B).
\section*{Discussion}
We have demonstrated an interaction-mediated plasmon splitter based on two counter-propagating QAH edge states without electron tunneling between the channels. A floating metallic island provides a common electrostatic environment for two spatially separated edges, giving rise to frequency-dependent redistribution of an injected plasmon between the two output channels. The observed oscillations in the splitting ratio are quantitatively reproduced by a long-range interaction model and provide direct access to the propagation and relaxation of the hybridized plasmon modes. A central consequence of the engineered electrostatic environment is the strong reduction of the plasmon velocity. We find $v_{\rm s}=(12\pm4)$~km\,s$^{-1}$ below the metallic island, compared with $(300-500)$~km\,s$^{-1}$ reported for isolated QAH edge states. The strongly reduced velocity of the propagation mode demonstrates that the metallic island provides an efficient means to engineer Coulomb interactions between chiral edge states.
The coupling geometry controls the interaction length and accumulated plasmon phase, providing a direct means to engineer the response of the splitter. This approach provides a route toward interaction-engineered chiral plasmonic circuits at zero magnetic field. Future control over the coupling geometry and electrostatic environment could enable coherent mode conversion, plasmon interferometry, and interaction-induced fractionalization in QAH systems.
The present device also highlights important limitations for future quantum applications. The observed approach toward a symmetric splitting ($\eta=1$) at high frequency is not a lossless beam-splitter operation, but results from dissipation in the plasmon splitter, which suppresses coherent oscillations and drives the system toward equal partitioning. Reducing these losses and extending the coherence of the collective plasmon modes will be essential for realizing fully coherent plasmonic quantum circuits. \newline\newline


\backmatter
\section*{Methods}
\subsection*{Fabrication}\label{par:Fabrication}
For the device fabrication, we use $8$~nm thin films of V-doped (Bi$_{1-x}$Sb$_x$)$_2$Te$_3$ grown via molecular beam epitaxy on InP (111)A substrates. The Bi:Sb ratio is optimized to place the chemical potential within the magnetic exchange gap without electrostatic gating, as confirmed by the observation of the QAH state in Hall measurements. The V-BST layer is protected by a $4$~nm-thick layer of $\text{Al}_2\text{O}_3$, which is deposited globally via low-temperature atomic layer deposition (ALD). The device fabrication follows the cleanroom recipes described in Ref.~\cite{roper_propagation_2024}. All patterning steps were performed using electron beam lithography with the e-beam resist PMMA A4. After patterning, the resist was developed using a mixture of methyl isobutyl ketone (MIBK) and IPA in a 1:3 ratio, followed by rinsing in IPA. Each lift-off of residual resist was performed using the solvent N-methyl-2-pyrrolidone (NMP). The fabrication comprises the following steps: First, ohmic contacts are established by wet etching the capping layer with Transene aluminum etchant type D and immediately depositing $5$~nm/$30$~nm of Pt/Au via sputtering afterwards. Second, the mesa was dry etched via Ar-sputtering, yielding two V-BST bars separated by a $1$~µm-wide trench. Third, a $10$~nm-thick layer of $\text{Al}_2\text{O}_3$ was globally deposited using ALD. It serves as the dielectric layer for the metallic island and the microwave finger gates. On top of the $\text{Al}_2\text{O}_3$ layer, the rectangular-shaped metallic island and the $3$~µm-wide finger gates are deposited by sputtering $5$~nm/$30$~nm of Pt/Au. Thus, the metallic island is capacitively coupled to the edge channels of both V-BST parts as indicated in Fig.~\ref{fig:TransmissionLine}B. The geometry of the finger gates is optimized to minimize direct microwave crosstalk, ensuring that the measured transmission predominantly probes propagation through the edge channels. In the final step, most of the $\text{Al}_2\text{O}_3$ layer was wet-etched such that it only covers the V-BST mesa and parts around it, and the coplanar waveguides and bonding pads were patterned using a $5$~nm/$45$~nm-thick layer of Pt/Au.
\subsection*{Measurement}\label{par:Measurement}
\begin{figure}[h]
        \centering
        \includegraphics[width=0.8\textwidth,keepaspectratio=true]{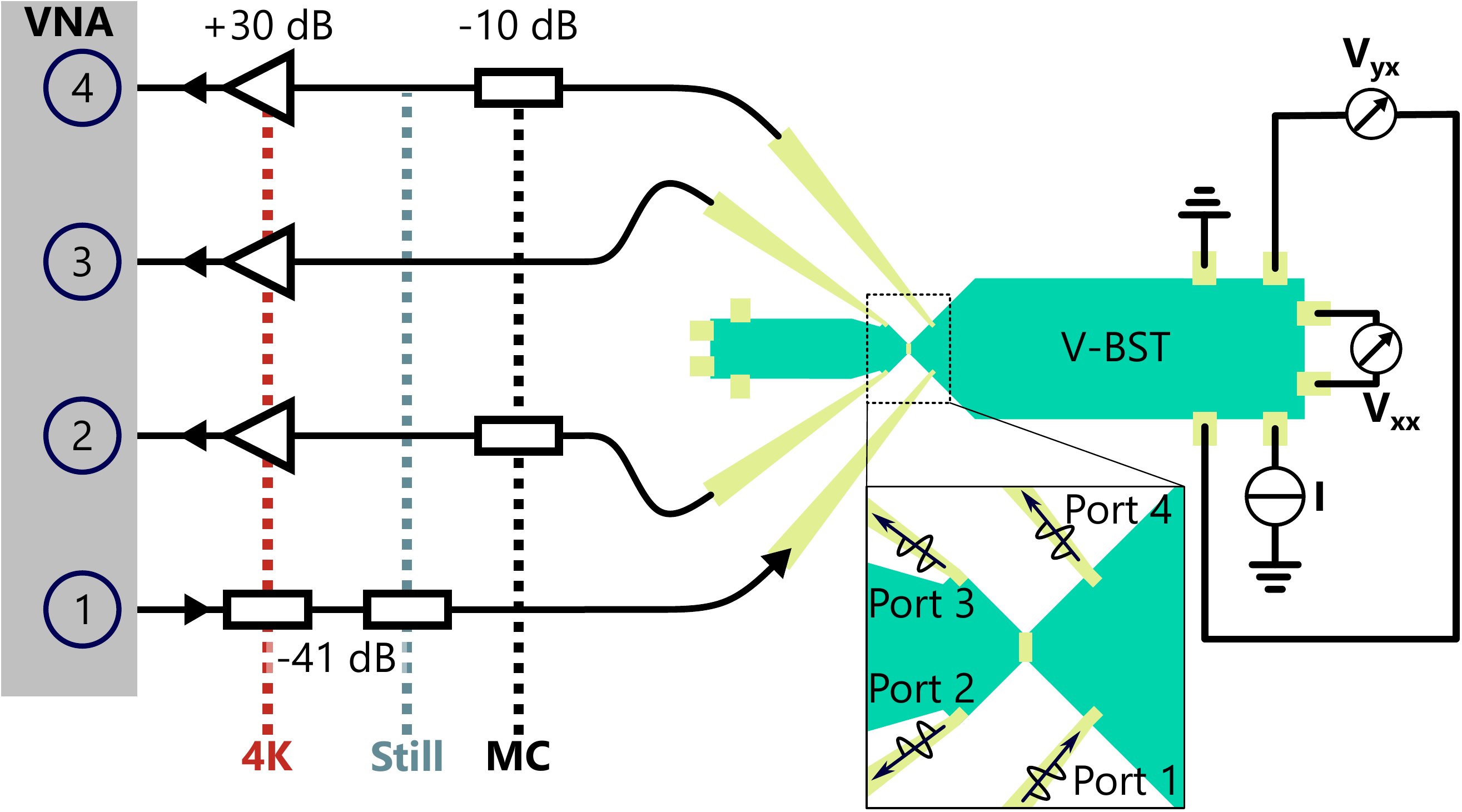}
        
        \caption{\textbf{DC and microwave setup.} A four-port VNA is connected to all four finger gates of the device, enabling measurement of the three transmission parameters $S_{21}$, $S_{31}$, and $S_{41}$. The input line (connected to port 1 of the VNA) contains attenuators of total -41~dB. Each of the detection lines is equipped with a cryogenic low noise amplifier (+30~dB). Detection lines 2 and 4 additionally comprise an attenuator (-10~dB) in order to dampen the standing wave forming between amplifier and sample. The dc wiring is exemplified for one V-BST bar on the right-hand side.}
        \label{fig:setup}
\end{figure}

The devices are measured in a Bluefors dilution refrigerator LD250 at a base temperature below $20$~mK. The fabricated devices are glued on a microwave-compatible PCB comprising 12 dc lines and 4 microwave lines connecting to the sample via coplanar waveguides. This allows us to characterize both V-BST bars during the same cooldown, as well as connecting and measuring all microwave ports simultaneously. A scheme of the setup and the components inside the dilution refrigerator is shown in Fig.~\ref{fig:setup}. Table \ref{tab:Device-list} lists the state of quantization assessed by the dc Hall measurements for each device.

\begin{table}[h!]
    \centering
    \captionsetup{width=\textwidth}
    \caption{\justifying\label{tab:Device-list} \textbf{Overview of samples.} Hall resistance $R_{\rm yx}$, longitudinal resistance $R_{\rm xx}$, and the coupling of all devices. All measurements are taken for $\mu_0H=2$~T and $T\approx 20$~mK.}
    \begin{tabular}{c|c|c|c}
    \textrm{Sample}&
    \textrm{$R_{\rm yx}/R_{\rm K}$~[\%]} & $R_{\rm xx}$~[$\Omega$] & Coupling \\ \hline
    A (Dec1A) &  $\mathrm{99.3 \pm 0.2}$ & $\mathrm{170 \pm 20}$ & 6~µm island \\
    B (Dec17B) &  $\mathrm{98.7 \pm 0.2}$ & $\mathrm{200 \pm 30}$ & 12~µm island \\
    C (Mar6B) &  $\mathrm{129 \pm 4}$ & $\mathrm{1200 \pm 50}$ & no island\\
    D (Jul9A)&  $\mathrm{96.9 \pm 0.3}$ & $\mathrm{600 \pm 70}$ & 6~µm island \\
    E (Mar6B) &  $\mathrm{129 \pm 4}$ & $\mathrm{1200 \pm 50}$ & 2~µm island \\
    F (Oct9A) &  $\mathrm{99.3 \pm 0.2}$ & $\mathrm{23 \pm 10}$ & no island\\
    \end{tabular}
\end{table}
\noindent We study microwave transport using a four-port vector network analyzer (VNA), which is connected via broadband coaxial cables to the wiring of the dilution refrigerator. It allows us to inject microwave signals into port 1, creating edge plasmons that propagate along the edge channels of the two V-BST bars. We then measure the transmission to the remaining three ports 2, 3, and 4 by relating the input signal of the VNA to its detected signal and obtain the transmission coefficients $S_{21}$, $S_{31}$, and $S_{41}$. In the main text, we focus on an input power of $P=-10$~dBm applied from the VNA. Inside the cryostat, the input signal is attenuated at all stages, leading to an attenuation of $-41$~dB. This corresponds to about $-51$~dBm on the excitation port. Additional attenuators ($-10$~dB) are placed on the output lines to dampen the standing wave forming between the device and the broadband amplifiers (see Ref.~\cite{roper_propagation_2024}).
To correct the signal for stray couplings, we apply the robust calibration scheme described in Ref.~\cite{roper_propagation_2024}, based on the following principle: for counterclockwise chirality (at $+2$~T), the edge plasmon will propagate along a long path away from the splitter and dissipate. In this case, the measured signal is merely the stray signal, which we subtract from the transmission for clockwise chirality (at $-2$~T). In the main text, we only show the calibrated transmission data.\newline\newline
To measure the longitudinal $R_{\rm xx}$ and transverse resistance $R_{\rm yx}$, we apply a current bias of $1$~nA and record the voltages $V_{\rm xx}$ and $V_{\rm yx}$ using standard lock-in techniques. Due to the geometry of the ohmic contacts, mixing of $R_{\rm xx}$ and $R_{\rm yx}$ occurs, which becomes visible when the bulk becomes conducting, e.g., as small peaks in the hysteresis loop of $R_{\rm yx}$ at the coercive field. The Hall resistance reaches $R_{\rm yx}=h/e^2$ within an accuracy of $2$~\%, while the longitudinal resistance remains below $R_{\rm xx}\lesssim1$~k$\Omega$. Devices C and E are an exception to that and exhibit a larger longitudinal resistance of $R_{\rm xx}\approx1.2$~k$\Omega$ and not a quantized Hall resistance $R_{\rm yx}\approx 1.29 h/e^2$. The hysteretic reversal of $R_{\rm yx}$ occurs at a coercive field of approximately $\mu_0 H_{\rm c}\approx1$~T. Table~\ref{tab:Device-list} in the method section lists the state of quantization assessed by the dc Hall measurements for each device.
\subsection*{Fitting procedure}

We determine the parameters of the transmission model in two successive fitting steps. First, we extract the parameters describing the plasmon splitter from the ratio of the two output signals, $S_{21}/S_{41}$. Since both transmission paths share the same excitation and propagation outside the coupling region, these contributions cancel, up to minor asymmetries in the device. We therefore fit the measured complex ratio using the splitter model of Eq.~2. For devices containing a metallic island, we work in the limit $\omega \tau_{\rm c,s}\gg1$ for which the dependence on the charging time $\tau_{\rm c,s}$ drops out. We therefore only use the propagation time $\tau_{\rm p,s}$ and relaxation time $\tau_{\rm r,s}$ as fit parameter. For the reference devices without a metallic island, $\tau_{\rm r,s}$ is set to zero and $\tau_{\rm c,s}$ is retained as a fit parameter to account for residual capacitive coupling between the two edge channels.\newline\newline
In a second step, the extracted parameters are kept fixed, and the complete transmission model of Eq.~3 is fitted simultaneously to the measured scattering parameters. This fit determines the parameters describing propagation along the bare edge channels, $\tau_{\rm p,e}$ and $\tau_{\rm r,e}$, as well as the gate capacitances $C_{\rm g}$ and the attenuation due to residual bulk conduction. The two output paths are fitted independently where required to account for differences in the excitation and detection gates.\newline\newline
For the fits, we simultaneously consider 
the in-phase and quadrature components and fit over the complete experimental frequency range using a nonlinear least-squares minimization provided by the {\sl scipy} package in Python. Parameter uncertainties are obtained from the covariance matrix of the fit. The same procedure is applied to all devices, allowing the extracted propagation and interaction parameters to be compared consistently between devices with and without a metallic island. For Device E, featuring a 2~µm-long island, parameter bounds were imposed (zero to infinity) to prevent the optimization from converging to negative values.

\section*{Data, code and materials availability}

The datasets generated and analyzed during the current study are available in
the Zenodo repository with the DOI \href{https://doi.org/10.5281/zenodo.22676232}{10.5281/zenodo.22676232}.


\begin{thebibliography}{10}
\providecommand{\url}[1]{{#1}}
\providecommand{\urlprefix}{URL }
\providecommand{\doi}[1]{\url{https://doi.org/#1}}




\bibitem{chang_experimental_2013}
C.Z. Chang, J.~Zhang, X.~Feng, J.~Shen, Z.~Zhang, M.~Guo, K.~Li, Y.~Ou, P.~Wei, L.L. Wang, Z.Q. Ji, Y.~Feng, S.~Ji, X.~Chen, J.~Jia, X.~Dai, Z.~Fang, S.C. Zhang, K.~He, Y.~Wang, L.~Lu, X.C. Ma, Q.K. Xue, Experimental {Observation} of the {Quantum} {Anomalous} {Hall} {Effect} in a {Magnetic} {Topological} {Insulator}.
\newblock Science \textbf{340}(6129), 167--170 (2013).
\newblock \doi{10.1126/science.1234414}.

\bibitem{chang_high-precision_2015}
C.Z. Chang, W.~Zhao, D.Y. Kim, H.~Zhang, B.A. Assaf, D.~Heiman, S.C. Zhang, C.~Liu, M.H.W. Chan, J.S. Moodera, High-precision realization of robust quantum anomalous {Hall} state in a hard ferromagnetic topological insulator.
\newblock Nature Materials \textbf{14}(5), 473--477 (2015).
\newblock \doi{10.1038/nmat4204}.

\bibitem{yu_quantized_2010}
R.~Yu, W.~Zhang, H.J. Zhang, S.C. Zhang, X.~Dai, Z.~Fang, Quantized {Anomalous} {Hall} {Effect} in {Magnetic} {Topological} {Insulators}.
\newblock Science \textbf{329}(5987), 61--64 (2010).
\newblock \doi{10.1126/science.1187485}.

\bibitem{roper_propagation_2024}
T.~Röper, H.~Thomas, D.~Rosenbach, A.~Uday, G.~Lippertz, A.~Denis, P.~Morfin, A.A. Taskin, Y.~Ando, E.~Bocquillon, Propagation, dissipation, and breakdown in quantum anomalous {Hall} edge states probed by microwave edge plasmons.
\newblock Physical Review B \textbf{110}(16), L161403 (2024).
\newblock \doi{10.1103/PhysRevB.110.L161403}.

\bibitem{mahoney_zero-field_2017}
A.C. Mahoney, J.I. Colless, L.~Peeters, S.J. Pauka, E.J. Fox, X.~Kou, L.~Pan, K.L. Wang, D.~Goldhaber-Gordon, D.J. Reilly, Zero-field edge plasmons in a magnetic topological insulator.
\newblock Nature Communications \textbf{8}(1), 1836 (2017).
\newblock \doi{10.1038/s41467-017-01984-5}.
\newblock Number: 1

\bibitem{martinez_edge_2023}
L.A. Martinez, G.~Qiu, P.~Deng, P.~Zhang, K.G. Ray, L.~Tai, M.T. Wei, H.~He, K.L. Wang, J.L. DuBois, D.X. Qu.
\newblock Edge {Magnetoplasmon} {Dispersion} and {Time}-{Resolved} {Plasmon} {Transport} in a {Quantum} {Anomalous} {Hall} {Insulator} (2023).
\newblock ArXiv:2308.15665 [cond-mat]

\bibitem{rupp_dynamics_2026}
A.~Rupp, D.~Rosenbach, T.~Röper, D.~Hoborka, A.A. Taskin, Y.~Ando, E.~Bocquillon, Dynamics of {Current}-{Induced} {Switching} in the {Quantum} {Anomalous} {Hall} {Effect}.
\newblock Physical Review Letters \textbf{136}(8), 086602 (2026).
\newblock \doi{10.1103/nlqq-w287}.

\bibitem{yuan_electrical_2024}
W.~Yuan, L.J. Zhou, K.~Yang, Y.F. Zhao, R.~Zhang, Z.~Yan, D.~Zhuo, R.~Mei, Y.~Wang, H.~Yi, M.H.W. Chan, M.~Kayyalha, C.X. Liu, C.Z. Chang, Electrical switching of the edge current chirality in quantum anomalous {Hall} insulators.
\newblock Nature Materials \textbf{23}(1), 58--64 (2024).
\newblock \doi{10.1038/s41563-023-01694-y}.

\bibitem{zimmermann_current-induced_2024}
E.~Zimmermann, J.~Teller, M.~Schleenvoigt, G.~Behner, P.~Schüffelgen, H.~Lüth, D.~Grützmacher, T.~Schäpers, Current-induced magnetization switching in a magnetic topological insulator heterostructure.
\newblock Physical Review Materials \textbf{8}(2), 026201 (2024).
\newblock \doi{10.1103/PhysRevMaterials.8.026201}.

\bibitem{viola_hall_2014}
G.~Viola, D.P. DiVincenzo, Hall {Effect} {Gyrators} and {Circulators}.
\newblock Physical Review X \textbf{4}(2), 021019 (2014).
\newblock \doi{10.1103/PhysRevX.4.021019}.

\bibitem{mahoney_-chip_2017}
A.C. Mahoney, J.I. Colless, S.J. Pauka, J.M. Hornibrook, J.D. Watson, G.C. Gardner, M.J. Manfra, A.C. Doherty, D.J. Reilly, On-chip microwave quantum hall circulator.
\newblock Physical Review X \textbf{7}(1), 011007 (2017)

\bibitem{rodenbach_metrological_2022}
L.K. Rodenbach, A.R. Panna, S.U. Payagala, I.T. Rosen, M.P. Andersen, P.~Zhang, L.~Tai, K.L. Wang, D.G. Jarrett, R.E. Elmquist, D.B. Newell, D.~Goldhaber-Gordon, A.F. Rigosi, Metrological {Assessment} of {Quantum} {Anomalous} {Hall} {Properties}.
\newblock Physical Review Applied \textbf{18}(3), 034008 (2022).
\newblock \doi{10.1103/PhysRevApplied.18.034008}.

\bibitem{huang_quantum_2025}
N.J. Huáng, J.L. Boland, K.M. Fijalkowski, C.~Gould, T.~Hesjedal, O.~Kazakova, S.~Kumar, H.~Scherer, Quantum anomalous {Hall} effect for metrology.
\newblock Applied Physics Letters \textbf{126}(4), 040501 (2025).
\newblock \doi{10.1063/5.0233689}.

\bibitem{beenakker_deterministic_2019}
C.~Beenakker, P.~Baireuther, Y.~Herasymenko, I.~Adagideli, L.~Wang, A.~Akhmerov, Deterministic creation and braiding of chiral edge vortices.
\newblock Physical Review Letters \textbf{122}(14), 146803 (2019).
\newblock \doi{10.1103/PhysRevLett.122.146803}.

\bibitem{giamarchi_quantum_2003}
T.~Giamarchi, \emph{Quantum {Physics} in {One} {Dimension}} (Oxford University Press, 2003).
\newblock \doi{10.1093/acprof:oso/9780198525004.001.0001}.

\bibitem{kumada_edge_2011}
N.~Kumada, H.~Kamata, T.~Fujisawa, Edge magnetoplasmon transport in gated and ungated quantum {Hall} systems.
\newblock Physical Review B \textbf{84}(4), 045314 (2011).
\newblock \doi{10.1103/PhysRevB.84.045314}.

\bibitem{kamata_voltage-controlled_2010}
H.~Kamata, T.~Ota, K.~Muraki, T.~Fujisawa, Voltage-controlled group velocity of edge magnetoplasmon in the quantum {Hall} regime.
\newblock Physical Review B \textbf{81}(8), 085329 (2010).
\newblock \doi{10.1103/PhysRevB.81.085329}.

\bibitem{bocquillon_separation_2013}
E.~Bocquillon, V.~Freulon, J.M. Berroir, P.~Degiovanni, B.~Plaçais, A.~Cavanna, Y.~Jin, G.~Fève, Separation of neutral and charge modes in one-dimensional chiral edge channels.
\newblock Nature Communications \textbf{4}(1), 1839 (2013).
\newblock \doi{10.1038/ncomms2788}.

\bibitem{levkivskyi_energy_2012}
I.P. Levkivskyi, E.V. Sukhorukov, Energy relaxation at quantum {Hall} edge.
\newblock Physical Review B \textbf{85}(7), 075309 (2012).
\newblock \doi{10.1103/PhysRevB.85.075309}.

\bibitem{berg_fractional_2009}
E.~Berg, Y.~Oreg, E.A. Kim, F.~Von~Oppen, Fractional {Charges} on an {Integer} {Quantum} {Hall} {Edge}.
\newblock Physical Review Letters \textbf{102}(23), 236402 (2009).
\newblock \doi{10.1103/PhysRevLett.102.236402}.

\bibitem{bocquillon_coherence_2013}
E.~Bocquillon, V.~Freulon, J.M. Berroir, P.~Degiovanni, B.~Plaçais, A.~Cavanna, Y.~Jin, G.~Fève, Coherence and {Indistinguishability} of {Single} {Electrons} {Emitted} by {Independent} {Sources}.
\newblock Science \textbf{339}(6123), 1054--1057 (2013).
\newblock \doi{10.1126/science.1232572}.

\bibitem{glidic_quasiparticle_2023}
P.~Glidic, O.~Maillet, C.~Piquard, A.~Aassime, A.~Cavanna, Y.~Jin, U.~Gennser, A.~Anthore, F.~Pierre, Quasiparticle {Andreev} scattering in the v = 1/3 fractional quantum {Hall} regime.
\newblock Nature Communications \textbf{14}(1), 514 (2023).
\newblock \doi{10.1038/s41467-023-36080-4}

\bibitem{frigerio_gate_2024}
E.~Frigerio, G.~Rebora, M.~Ruelle, H.~Souquet-Basiège, Y.~Jin, U.~Gennser, A.~Cavanna, B.~Plaçais, E.~Baudin, J.M. Berroir, I.~Safi, P.~Degiovanni, G.~Fève, G.C. Ménard, Gate tunable edge magnetoplasmon resonators.
\newblock Communications Physics \textbf{7}(1), 314 (2024).
\newblock \doi{10.1038/s42005-024-01803-6}.

\bibitem{kouame_electrostatic_2026}
S.~Kouamé, E.~Frigerio, G.~Rebora, S.~Purkait, Y.~Jin, U.~Gennser, A.~Cavanna, J.M. Berroir, E.~Baudin, P.~Degiovanni, F.D. Parmentier, G.~Fève, G.C. Ménard.
\newblock Electrostatic splitting of an {Edge} {Magnetoplasmon} {Resonator} (2026).
\newblock \doi{10.48550/arXiv.2609.05188}.
\newblock ArXiv:2609.05188 [cond-mat.mes-hall]

\bibitem{hashisaka_distributed-element_2013}
M.~Hashisaka, H.~Kamata, N.~Kumada, K.~Washio, R.~Murata, K.~Muraki, T.~Fujisawa, Distributed-element circuit model of edge magnetoplasmon transport.
\newblock Physical Review B \textbf{88}(23), 235409 (2013).
\newblock \doi{10.1103/PhysRevB.88.235409}.

\bibitem{kamata_fractionalized_2014}
H.~Kamata, N.~Kumada, M.~Hashisaka, K.~Muraki, T.~Fujisawa, Fractionalized wave packets from an artificial {Tomonaga}–{Luttinger} liquid.
\newblock Nature Nanotechnology \textbf{9}(3), 177--181 (2014).
\newblock \doi{10.1038/nnano.2013.312}.

\bibitem{perfetto_time-resolved_2014}
E.~Perfetto, G.~Stefanucci, H.~Kamata, T.~Fujisawa, Time-resolved charge fractionalization in inhomogeneous {Luttinger} liquids.
\newblock Physical Review B \textbf{89}(20), 201413 (2014).
\newblock \doi{10.1103/PhysRevB.89.201413}.

\bibitem{martinez_edge_2024}
L.A. Martinez, G.~Qiu, P.~Deng, P.~Zhang, K.G. Ray, L.~Tai, M.T. Wei, H.~He, K.L. Wang, J.L. DuBois, D.X. Qu, Edge magnetoplasmon dispersion and time-resolved plasmon transport in a quantum anomalous {Hall} insulator.
\newblock Physical Review Research \textbf{6}(1), 013081 (2024).
\newblock \doi{10.1103/PhysRevResearch.6.013081}.

\bibitem{roper_breakdown_2025}
T.~Röper, D.~Rosenbach, A.~Rosch, A.A. Taskin, Y.~Ando, E.~Bocquillon, Breakdown of the quantum anomalous {Hall} effect under microwave drives.
\newblock Physical Review B \textbf{112}(20), L201301 (2025).
\newblock \doi{10.1103/3148-3vp4}.

\bibitem{gourmelon_characterization_2020}
A.~Gourmelon, H.~Kamata, J.M. Berroir, G.~Fève, B.~Plaçais, E.~Bocquillon, Characterization of helical {Luttinger} liquids in microwave stepped-impedance edge resonators.
\newblock Physical Review Research \textbf{2}(4), 043383 (2020).
\newblock \doi{10.1103/PhysRevResearch.2.043383}.

\bibitem{lee_spin-charge_1997}
H.C. Lee, S.R. Eric~Yang, Spin-charge separation in quantum {Hall} edge liquids.
\newblock Physical Review B \textbf{56}(24), R15529--R15532 (1997).
\newblock \doi{10.1103/PhysRevB.56.R15529}.

\bibitem{sukhorukov_resonant_2007}
E.V. Sukhorukov, V.V. Cheianov, Resonant {Dephasing} in the {Electronic} {Mach}-{Zehnder} {Interferometer}.
\newblock Physical Review Letters \textbf{99}(15), 156801 (2007).
\newblock \doi{10.1103/PhysRevLett.99.156801}.

\bibitem{kumada_suppression_2020}
N.~Kumada, N.H. Tu, K.i. Sasaki, T.~Ota, M.~Hashisaka, S.~Sasaki, K.~Onomitsu, K.~Muraki, Suppression of gate screening on edge magnetoplasmons by highly resistive {ZnO} gate.
\newblock Physical Review B \textbf{101}(20), 205205 (2020).
\newblock \doi{10.1103/PhysRevB.101.205205}.

\bibitem{flensberg_coulomb_1998}
K.~Flensberg, Coulomb {Drag} of {Luttinger} {Liquids} and {Quantum} {Hall} {Edges}.
\newblock Physical Review Letters \textbf{81}(1), 184--187 (1998).
\newblock \doi{10.1103/PhysRevLett.81.184}.

\bibitem{fu_non-reciprocal_2025}
Y.~Fu, Y.~Huang, Q.L. He, Non-reciprocal {Coulomb} drag between {Chern} insulators.
\newblock Nature Communications \textbf{16}(1), 3058 (2025).
\newblock \doi{10.1038/s41467-025-58401-5}.

\end{thebibliography}

\section*{Acknowledgments}

\subsection*{Funding:}
This work has been supported by the Germany Excellence Strategy (Cluster of Excellence Matter and Light for Quantum Computing ML4Q, EXC 2004/1 - 390534769) and the DFG (SFB1238 Control and Dynamics of Quantum Materials, 277146847, projects A04 and B07).
\subsection*{Author contributions:}
D.H. fabricated the devices, and A.A.T. grew the films. D.H. and T.R. performed the measurements and analyzed the data. F.H. and E.B. assisted with the data analysis and interpretation. D.H. and T.R. prepared the manuscript. E.B. supervised the project. All authors discussed the results and contributed to the writing of the manuscript.
\subsection*{Competing interests:}
All authors declare no financial or non-financial competing interests.
\section*{Supplementary information}
The article contains supplementary material available at [insert URL later]

\includepdf[pages=-]{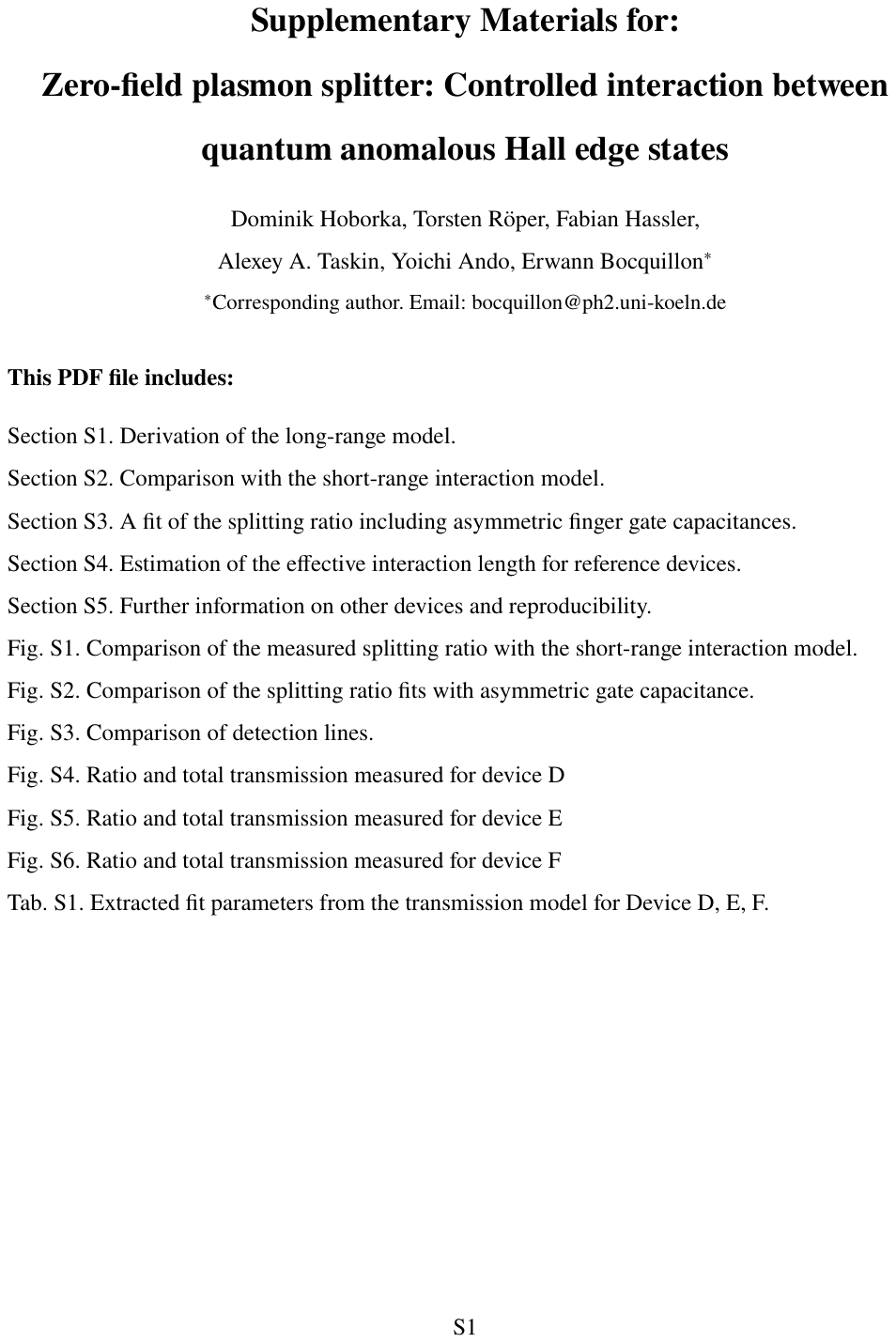}
\end{document}